\documentclass[journal]{IEEEtran}
\ifCLASSINFOpdf
\else
\fi

\usepackage{amsthm}
\usepackage[cmex10]{amsmath}
\usepackage{amsfonts}
\usepackage{amssymb}

\ifCLASSINFOpdf
\usepackage[pdftex]{graphicx}
\usepackage{subfigure}
\else

\usepackage{graphicx}
\usepackage{subfigure}
\fi
\usepackage{epstopdf}
\usepackage{color,xcolor}
\usepackage{multicol}  
\usepackage{multirow}
\usepackage{booktabs} 
\usepackage{threeparttable}
\usepackage{algorithmic,algorithm}
\usepackage{subfigure}
\usepackage{tabularx}
\usepackage{enumitem}

\usepackage[font=normalsize]{caption}

\usepackage{algorithmic,algorithm}
\renewcommand{\algorithmicrequire}{\textbf{Input:}}
\renewcommand{\algorithmicensure}{\textbf{Output:}}

\renewcommand{\algorithmicrequire}{\textbf{Input:}}
\renewcommand{\algorithmicensure}{\textbf{Output:}}

\usepackage[numbers,sort&compress]{natbib}
\usepackage{color}

\usepackage{url}

\usepackage{mdwtab}

\usepackage{hyperref} % reference link

\begin{document}
%
% paper title
% Titles are generally capitalized except for words such as a, an, and, as,
% at, but, by, for, in, nor, of, on, or, the, to and up, which are usually
% not capitalized unless they are the first or last word of the title.
% Linebreaks \\ can be used within to get better formatting as desired.
% Do not put math or special symbols in the title.
\title{FedEdge AI-TC: A Semi-supervised Traffic Classification Method based on Trusted Federated Deep Learning for Mobile Edge Computing}
%
%
% author names and IEEE memberships
% note positions of commas and nonbreaking spaces ( ~ ) LaTeX will not break
% a structure at a ~ so this keeps an author's name from being broken across
% two lines.
% use \thanks{} to gain access to the first footnote area
% a separate \thanks must be used for each paragraph as LaTeX2e's \thanks
% was not built to handle multiple paragraphs
%

\author{Pan~Wang,~\IEEEmembership{Member,~IEEE,}
        Zeyi~Li, Mengyi~Fu, Zixuan~Wang, Ze~Zhang  and~Minyao~Liu% <-this % stops a space
\thanks{Pan Wang was with the School
of Modern Posts, Nanjing University of Post\&Telecommunications, Nanjing,
Jiangsu, 210003 China e-mail: (wangpan@njupt.edu.cn).}% <-this % stops a space
\thanks{Manuscript received July 20, 2023.}}

% note the % following the last \IEEEmembership and also \thanks - 
% these prevent an unwanted space from occurring between the last author name
% and the end of the author line. i.e., if you had this:
% 
% \author{....lastname \thanks{...} \thanks{...} }
%                     ^------------^------------^----Do not want these spaces!
%
% a space would be appended to the last name and could cause every name on that
% line to be shifted left slightly. This is one of those "LaTeX things". For
% instance, "\textbf{A} \textbf{B}" will typeset as "A B" not "AB". To get
% "AB" then you have to do: "\textbf{A}\textbf{B}"
% \thanks is no different in this regard, so shield the last } of each \thanks
% that ends a line with a % and do not let a space in before the next \thanks.
% Spaces after \IEEEmembership other than the last one are OK (and needed) as
% you are supposed to have spaces between the names. For what it is worth,
% this is a minor point as most people would not even notice if the said evil
% space somehow managed to creep in.

% The paper headers
\markboth{Journal of \LaTeX\ Class Files,~Vol.~14, No.~8, August~2015}%
{Shell \MakeLowercase{\textit{et al.}}: Bare Demo of IEEEtran.cls for IEEE Journals}
% The only time the second header will appear is for the odd numbered pages
% after the title page when using the twoside option.
% 
% *** Note that you probably will NOT want to include the author's ***
% *** name in the headers of peer review papers.                   ***
% You can use \ifCLASSOPTIONpeerreview for conditional compilation here if
% you desire.

% If you want to put a publisher's ID mark on the page you can do it like
% this:
%\IEEEpubid{0000--0000/00\$00.00~\copyright~2015 IEEE}
% Remember, if you use this you must call \IEEEpubidadjcol in the second
% column for its text to clear the IEEEpubid mark.

% use for special paper notices
%\IEEEspecialpapernotice{(Invited Paper)}

% make the title area
\maketitle

% As a general rule, do not put math, special symbols or citations
% in the abstract or keywords.
\begin{abstract}
As a typical entity of MEC (Mobile Edge Computing), 5G CPE (Customer Premise Equipment) has proven to be a promising alternative to traditional HGU (Home Gateway Unit). Network TC (Traffic Classification) is a vital service quality assurance and security management method for communication networks, which has become a crucial functional entity in 5G CPE/HGU. In recent years, many researchers have applied Machine Learning (ML) or Deep Learning (DL) to TC, namely AI-TC, to improve its performance. However, AI-TC methods face significant challenges, including high data dependency, exhaustively costly traffic labeling, and network subscribers' privacy. Besides, as the AI-TC carrier, 5G CPE/HGU's limited computing resources often become the bottleneck of models for efficient classification. Furthermore, the long-standing problem of the "black box" for AI-TC models has always perplexed network operators regarding the model's transparency and credibility, i.e., AI model interpretability. Therefore, how to achieve an efficient and trusted classification carried on the "weak computing power" network entity while protecting user privacy has become the key to ensuring the service quality and security of the home network. This paper presents the FedEdge AI-TC framework, a novel AI-TC approach for implementing trusted Federated Learning (FL) based efficient Network TC in 5G CPE/HGU. First, FedEdge AI-TC effectively protects the data privacy of network subscribers by proposing an FL based framework of local training, model parameters iterating, and centralized training. Second, a semi-supervised TC algorithm based on Variational Auto-Encoder (VAE) and convolutional neural network (CNN) is designed to reduce data dependence while keeping the TC accuracy. Finally, XAI-Pruning, an AI model compression method, combined with the DL model interpretability, is proposed to condense the model and interpret it globally and locally to achieve light-weighted AI-TC model deployment while building the trust in their decision of network operators. To demonstrate the efficiency of the proposed method, we conducted some experimental evaluations on commonly used public benchmark datasets and real network datasets. The results show that FedEdge AI-TC can outperform the benchmarking methods regarding the accuracy and achieve excellent TC performance of model inference on 5G CPE/HGU with limited computing resources, which effectively protects the users' privacy and improve the model's credibility.
\end{abstract}

% Note that keywords are not normally used for peerreview papers.

% For peer review papers, you can put extra information on the cover
% page as needed:
% \ifCLASSOPTIONpeerreview
% \begin{center} \bfseries EDICS Category: 3-BBND \end{center}
% \fi
%
% For peerreview papers, this IEEEtran command inserts a page break and
% creates the second title. It will be ignored for other modes.
\IEEEpeerreviewmaketitle

\section{Introduction}\label{sec:intro}
\begin{figure}[htbp]
	\centerline{\includegraphics[width=9cm, height=4cm]{ 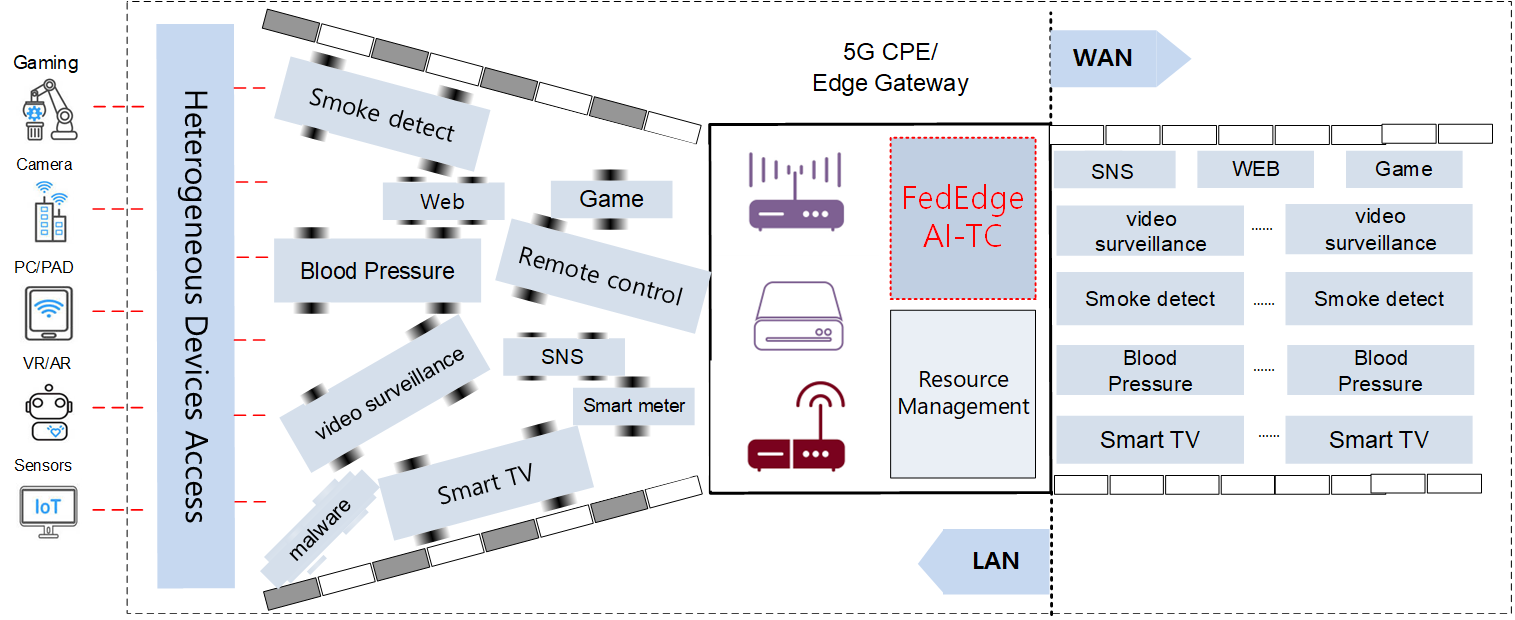}}
	\caption{The Scenario of AI-TC for MEC.}
	\label{fig_scenario}
\end{figure}

% The very first letter is a 2 line initial drop letter followed
% by the rest of the first word in caps.
% 
% form to use if the first word consists of a single letter:
% \IEEEPARstart{A}{demo} file is ....
% 
% form to use if you need the single drop letter followed by
% normal text (unknown if ever used by the IEEE):
% \IEEEPARstart{A}{}demo file is ....
% 
% Some journals put the first two words in caps:
% \IEEEPARstart{T}{his demo} file is ....
% 
% Here we have the typical use of a "T" for an initial drop letter
% and "HIS" in caps to complete the first word.
\IEEEPARstart{A}{s} a distinct entity of MEC, 5G CPE has gradually become an alternative to HGU. Network traffic classification has played a crucial role in ensuring service quality and managing security for home networks. It is a critical functional element within 5G CPE. It finds extensive applications in QoS (Quality of Service) / QoE (Quality of Experience) management, network resource optimization, congestion control, and intrusion detection. With the popularity of smart homes, many applications such as video surveillance, fire and smoke detection, smart appliances, VR/AR, and others have emerged alongside traditional internet services like high-definition videos and online gaming. These applications impose demanding requirements on the network's QoS, including fast and flexible customization of services, real-time responsiveness, and high reliability. Thus, home networks exhibit four significant trends: "Terminals Heterogeneity, Applications Diversity, High Privacy, and Rapid Evolution." Traffic classification in home networks, as an important prerequisite for fine-grained network resource management, has become one of the crucial security measures for smart homes. As shown in Fig.~\ref{fig_scenario}, the 5G CPE/Edge Gateway serves as the "connection point" between the smart home and the wide area network. It is crucial for the reliable forwarding of household application traffic. The AI-TC based on 5G CPE/Edge Gateway is the key to achieving fine-grained network resource management, QoE assurance, and intrusion detection in home networks.

The development of Network TC has generally gone through three stages. In the first phase, TC methods were mainly based on port matching or DPI (Deep Packet Inspection). However, this type of technology quickly became ineffective with the increasing use of techniques like tunneling, encryption, random ports, and concerns about security issues such as user privacy breaches. The second stage primarily leveraged machine learning techniques to extract underlying patterns of different services/applications/attacks traffic features and achieved TC by discriminating various applications in the data space. However, such methods require the extraction of high-quality traffic features as the training inputs for ML. The extraction and selection of these features heavily rely on the domain expertise of network specialists and are time-consuming and labor-intensive. In the third stage, with the rapid development of cloud computing, big data, especially deep learning, and high-performance computing technology, feature learning of massive traffic data has become feasible, bringing new imagination space for improving the TC's performance. DL has three excellent features: automatic feature extraction, exploration of deep nonlinear features, and many classical models  in computer vision/image/text/speech that can be reused. These advantages are all lacking in ML-based TC methods. Several DL-based TC technologies have been proposed recently, including CNN/AE/MLP/LSTM/GAN-based methods, which have achieved better classification performance than ML-TC\cite{1,2,3,4,5}.

However, applying DL technology to smart home network traffic classification faces three major challenges. Firstly, DL models heavily rely on a large volume of online behavior data from home users, which raises concerns about highly sensitive user privacy. Additionally, the collection and labeling of traffic samples are time-consuming and labor-intensive. Secondly, the 5G CPE/edge gateway's limited computing resources often become the bottleneck of the AI-TC models for efficient classification. Thirdly, the "black box" problem of the DL classification model has always perplexed the trustworthiness of home users/network operators. Therefore, efficiently achieving trusted classification of home network traffic on a "weak computing power" gateway device while protecting user privacy is crucial in ensuring service quality and security.

Federated Learning is a distributed machine learning technology providing privacy protection, presenting a novel application paradigm that balances data privacy protection and sharing computing\cite{9}. FL constructs a global model based on virtual fusion data through distributed model training among multiple data sources with local data, without exchanging local data but model parameters or intermediate results. In recent years, FL has been widely used in industries with high sensitivity to data privacy, such as finance and medical care, and has made significant progress\cite{10}. Inspired by this, this paper proposes FedEdge AI-TC, an AI traffic classification method based on federated learning of 5G CPE/edge gateway. This method uses the FL framework to train the traffic classification model of the home network without uploading home network data to a centralized server but executing local distributed model training on a 5G CPE/edge gateway. The global traffic classification model is constructed by exchanging model parameters with the centralized server while protecting the privacy of home users. In addition, considering that traffic sample collection and labeling are time-consuming and labor-intensive, we design a semi-supervised traffic classification algorithm based on VAE and CNN to reduce dependency on traffic sample data. Finally, XAI-Pruning, an AI model compression method, combined with the DL model interpretability, is proposed to condense the model and interpret it globally and locally to achieve light-weighted AI-TC model deployment while building the trust in their decision of network operators. To demonstrate the efficiency of the proposed method, we conducted some experimental evaluations on commonly used public benchmark datasets and real network datasets. The results show that FedEdge AI-TC can outperform the benchmarking methods regarding the accuracy and achieve excellent TC performance of model inference on 5G CPE/HGU with limited computing resources, which effectively protects the users' privacy and improve the model's credibility. The contributions of this paper are as follows:

\begin{enumerate}
    \item We propose a 5G CPE traffic classification method FedEdge AI-TC based on federated learning, which effectively protects the privacy of home user data by constructing the FL framework of local training, parameter updating, and centralized training;
    \item A semi-supervised traffic classification algorithm based on VAE and CNN is designed to reduce the dependence on traffic sample data;
    \item A pruning method based on DL model interpretability (XAI-Pruning) is proposed for model compression, and the model is globally and locally explained to increase model transparency and credibility;
    \item Experiments on public and self-built datasets show this method can achieve high traffic classification accuracy under limited computing resources.
\end{enumerate}

The chapter organization of this paper is as follows: Section \ref{sec:intro} is an overall introduction; Section \ref{sec:related_works} is related research works; Section \ref{sec:framework} presents the framework for the proposed approach; Section \ref{sec:method} describes the FedEdge AI-TC method; Section \ref{sec:evaluation} evaluates the proposed method through experiments and provides a comprehensive discussion of the results.; Section \ref{sec:conclusion} concludes the contributions and outlines potential directions for future research. Table \ref{table_abbreviations} below is the list of abbreviations in alphabetical order.

\begin{table}[htbp]
\centering
\caption{List of abbreviations in alphabetical order}
\label{table_abbreviations}
\begin{tabular}{ll}
		\hline \text { Acronym } & \text { Explanation } \\
		\hline \text { AE } & \text { Auto Encoders } \\
        \text { CPE } & \text { Customer Premise Equipment } \\
		\text { CNN } & \text { Convolutional Neural Networks } \\
		\text { DL } & \text { Deep Learning } \\
        \text { DPI } & \text { Deep Packet Inspection } \\
        \text { FL } & \text { Federated Learning } \\
        \text { FAM } & \text { Flow Attribute Matrix } \\
        \text { GAN } & \text { Generative Adversarial Network }\\
        \text { HGU } & \text { Home Gateway Unit } \\
        \text { LSTM } & \text { Long Short Term Memory }\\
        \text { MSE } & \text { Mean Square Error } \\
        \text { MLP } & \text { Multilayer Perceptron }\\
        \text { QoE } & \text { Quality of Experience }\\
        \text { QoS } & \text { Quality of Service }\\
        \text { SSL } & \text { Semi-Supervised Learning } \\
        \text { TC } & \text { Traffic Classification } \\
        \text { VAE } & \text { Variational Auto Encoder } \\
        \text { XAI } & \text { Model Explanation/Interpretability } \\
		\text { ML } & \text { Machine Learning } \\
        \text { GNN } & \text { Graph Neural Network } \\
		\hline
\end{tabular}
\end{table}

\section{Related works}\label{sec:related_works}
\subsection{Deep Learning based Traffic Classification}
Deep learning, also referred to as deep structured learning or hierarchical learning, is achieved by acquiring the representation of data. In contrast to traditional machine learning algorithms, deep learning can automatically extract features without human intervention, rendering it an ideal approach for traffic classification. The application of deep learning techniques to Network TC\cite{8,11,12,13,6} involves three steps: firstly, characterizing the input data by defining and designing the model input using data packets, PCAP files, or traffic statistics vectors as features; secondly, selecting suitable models and algorithms based on classifier objectives and model characteristics; finally extracting traffic features automatically through training a DL-based classifier and associating input data with corresponding category labels.

Recent research has demonstrated deep learning methods' superiority in traffic classification. For example, CNNs\cite{14,15} are widely used in traffic classification. They can automatically extract features from raw network traffic data and have end-to-end learning capabilities during training. In addition, RNNs and LSTMs\cite{16,17} are used to process traffic sequence data and can capture the temporal dependencies in the data. These types of networks are often used in traffic classification to identify persistent attacks such as DDoS (Distributed Denial of Service) attacks. GNN has proven to be a novel information representation method for DL, which has been applied in TC or IDS\cite{41}.

\subsection{Semi-supervised Learning based Traffic Classification}
Semi-supervised traffic classification\cite{18,19,20,21} is an approach that utilizes a small set of labeled data along with a substantial amount of unlabeled data to distinguish various network traffic types. There are four primary methods for semi-supervised traffic classification: cluster-based methods, generative models, GANs (Generative Adversarial Networks), and discriminative models. Cluster-based methods\cite{22,23} have low computational complexity but can be influenced by data distribution and may exhibit instability in practical usage. Generative model-based methods\cite{24} are effective for unknown or dynamically changing network applications; however, they require prior knowledge to select appropriate statistical features and clustering parameters which might limit the generalization of classification results. GAN-based methods\cite{25,7} can enhance dataset diversity and quality, thereby improving the model's generalization performance. Nevertheless, GAN models have complex structures and numerous parameters that pose challenges in training and render them impractical for deployment on edge devices. Therefore, this study primarily adopts a discriminative model-based approach by directly learning the mapping function from feature space to class space. Model parameters are optimized by minimizing the classification error of labeled data while incorporating a regularization term for unlabeled data. Subsequently, the unlabeled data is classified based on predicted results.

\begin{figure*}[ht]
	\centering\includegraphics[scale=0.5]{ 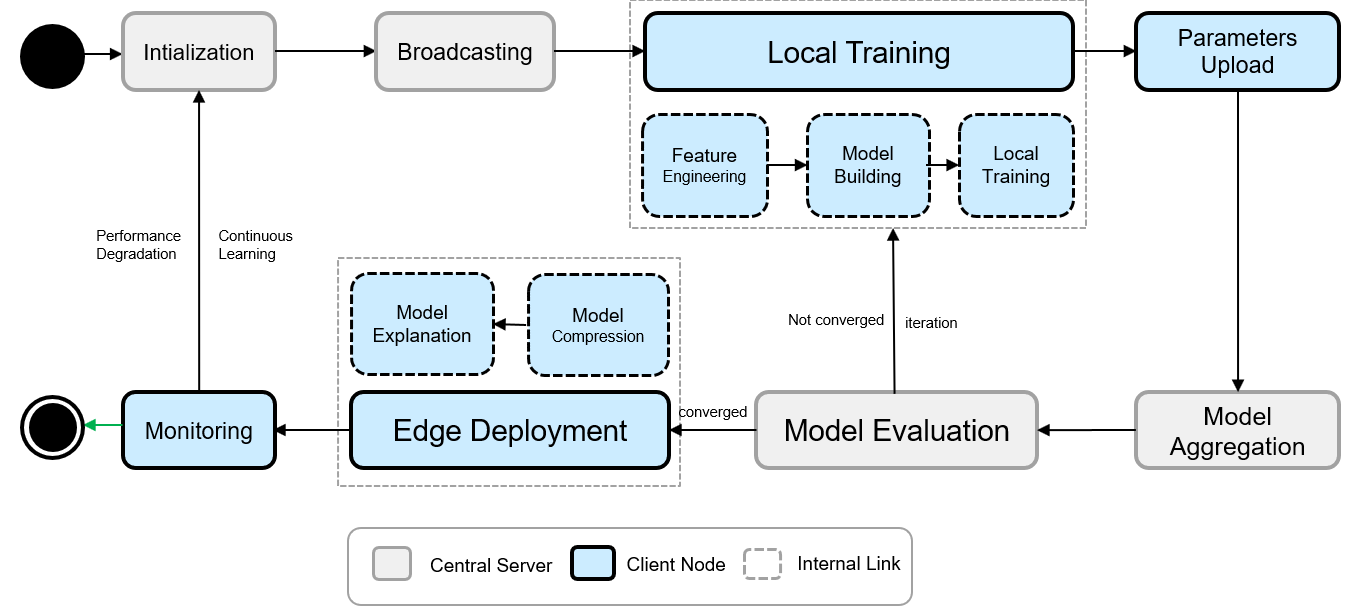}
	\caption{The Workflow of the End-to-End Traffic Classification based on Trusted Federated Learning for Edge Gateway.}
	\label{fig_workflow} % add a label for reference
\end{figure*}

\subsection{FL and its applications in Traffic Classification}
Federated Learning (FL) is a distributed ML/DL framework that focuses on decentralized training data, aiming to obtain ML/DL models by distributing the data across numerous nodes while ensuring privacy and security. FL allows local clients to retain their data, sharing only the model parameters with a central server, thereby reducing communication overhead and preserving client data privacy\cite{10,27}. Recently, two main approaches have emerged for traffic classification tasks by combining federated learning with deep learning techniques. The first approach\cite{28,29,30,43} empowers child nodes to annotate the data through various means. The second approach\cite{31,32,33,35,42} involves transforming the model structure and training objectives so that sub-nodes can train the model using unlabeled data; subsequently, fine-tuning is performed by the server using labeled data to achieve semi-supervised traffic classification. We propose a semi-supervised traffic classification model based on VAE-CNN that incorporates a federated learning paradigm enabling edge devices' semi-supervised training of the model.

\section{The Overall Framework}\label{sec:framework}
\subsection{The Workflow of FedEdge AI-TC}\label{sec:workflow}

Smart home networks face four major challenges: terminal heterogeneity, application diversity, high privacy, and rapid evolution. A network TC system must continuously learn through long-term iterative optimization to overcome these challenges. The process follows the full life cycle of federated learning, as shown in Fig.~\ref{fig_workflow}. The edge-side AI-TC classification system workflow based on federated learning includes \textbf{\emph{initialization, broadcast, training, parameter uploading, model aggregation and evaluation, edge deployment, and model monitoring}}. Here are the steps for implementing an AI-powered network traffic classification system:

\begin{enumerate}
    \item \textbf{Initialization:} Provide the client node with an initialization model for efficient local/global model training.
    
    \item \textbf{Broadcast:} The centralized server broadcasts the initialization model to all the client nodes like 5G CPE/HGU.
    
    \item \textbf{Local Training:} The client node performs feature engineering, model construction, and local training.
    
    \begin{enumerate}
        \item Feature Engineering: Extract, select, represent, and compress network traffic features to build an optimal feature subset for the AI-TC classification system.
        
        \item Model Construction and Local Training: This step involves selecting what kind of learning methods (supervised/semi-supervised/unsupervised/weakly supervised), training methods (centralized/distributed training, federated learning), whether to pre-train, whether to use classical models for transfer learning, to form a local model (Local Model) on the local client node.
    \end{enumerate}
    
    \item \textbf{Parameters Uploading:} Upload encrypted parameter information obtained from local training to the centralized server.
    
    \item \textbf{Aggregation:} The centralized server performs 'secure aggregation' on the parameter information uploaded by each client node (such as using the \emph{\textbf{FedAvg}} algorithm) and performs global training.
    
    \item \textbf{Model Evaluation:} Evaluate the global model obtained from the centralized server's global training. If the training process converges, it will enter the model deployment step; otherwise, it will inform the client node to continue training and iterative optimization. In addition, the model evaluation also needs to consider its computational complexity, time complexity, and the computing resources and time required for training/inference.
    
    \item \textbf{Edge deployment:}  Deploy the model on the edge or terminal side using the pull/push/subscribe model deploy method and update strategy. Model compression and interpretation are two important tasks in this step.
    \begin{enumerate}
    
        \item \textbf{Model Compression:} Compress the inference/classification model small enough to meet the fast classification under limited computing power. 
        
        \item \textbf{XAI(Model Interpretation/Explanation):} Solve the "black box" problem of the AI-TC model to make the classification model users trust the model.
        
    \end{enumerate}
    
    \item \textbf{Model Monitoring:}  Monitor the status of the classification system, including model, and real-time network flow, to report some key issues such as classification system failures and model degradation.
    
    \item \textbf{Continuous Learning:} Initiate iterative optimizations and continuous learning from initialization to maintain high adaptability, robustness, and reliability of the classification system.
\end{enumerate}

The following sections of this article will focus on \textbf{\emph{initialization, model training, compression, and explanation.}}

\subsection{The Architecture of FedEdge AI-TC}\label{sec:architecture}

\begin{figure*}[ht]
	\centering\includegraphics[scale=0.6]{ 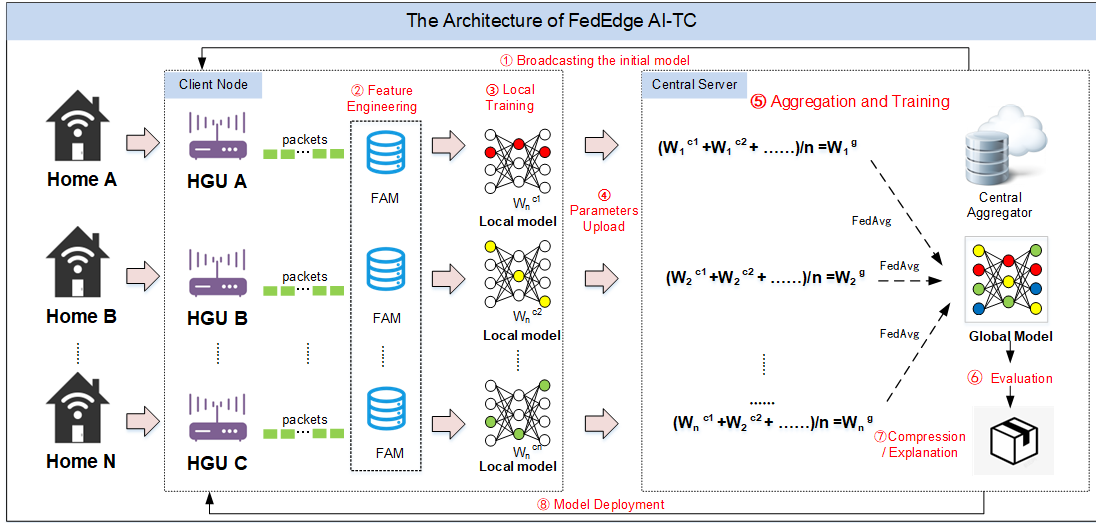}
	\caption{The Architecture of FedEdge AI-TC.}
	\label{fig_architecture} % add a label for reference
\end{figure*}

The overall architecture of FedEdge AI-TC is illustrated in Fig.~\ref{fig_architecture}, which is divided into the client node and central server/central aggregator based on our previous work\cite{45}. HGU, functioning as the client node, performs local training and inference classification tasks\cite{44}. On the other hand, the central aggregator acts as a centralized server responsible for aggregated training, model evaluation, compression, interpretation, and deployment of inference models. The workflow can be summarized as follows: Initially, the central aggregator broadcasts the initial classification model to HGU. Subsequently, HGU collects real-time packets and performs redundant/invalid packet filtering, network flow attribute calculation, and normalization to formulate a FAM (Flow Attribute Matrix), depicted in Fig.\ref{fig_FAM}. This matrix is then utilized for training an initialization model locally. Afterward, encrypted gradient information, loss, and other parameter details are uploaded to the central aggregator for aggregate training. Since each HGU exhibits similar flow characteristics, Horizontal Federated Learning (HFL) is employed for aggregate training by standard secure aggregation algorithms like FedAvg. Then the model performance metrics will be evaluated, including accuracy, precision, recall, and F1-Score for assessing model convergence. If convergence occurs successfully, the aggregated global model undergoes compression to meet computing resource constraints of HGU (including CPU/memory/Flash). Once an available inference model is obtained, XAI-based methods are applied to provide both global and local explanations of the model. Finally, the resulting inference model will be distributed across all HGUs. Otherwise, if convergence fails, the global model will be issued to each HGU, initiating a new epoch of iterative training and optimization until it converges.

\section{The Methodology of FedEdge AI-TC}\label{sec:method}
\subsection{Initial Model}
The initial model plays a crucial role in determining the convergence speed and performance of federated aggregation training, which serves as the initial stage of federated learning. While randomly initialized models can be used within the FL framework, it is essential to construct an initial model that enhances local/global model training efficiency. Unlike traditional AI domains like images and text, network traffic classification lacks pre-trained models, necessitating the development of the initial model by ourselves. In this study, we utilize three benchmarking datasets including (ISCIX\cite{37}, UNSW-NB15\cite{38}, and MIRAGE\cite{39}) as baseline datasets for constructing our initial model. We adopt CNN as the supervised training algorithm, and further details about the initial model can be found in our previous work\cite{45}.

\subsection{Federated Semi-Supervised Learning Traffic Classification Method using VAE+CNN}
\subsubsection{\textbf{The Introduction of FSSL}}
There are two types of network traffic data in the FedEdge AI-TC system. The first type, labeled data, is stored in the central aggregator. The second type, unlabeled data, is located in the local HGU, i.e., real-time traffic data. From the perspective of FL, it belongs to the disjoint scenario. 
This alignment with the home network scenario arises due to the absence of labels for real-time traffic forwarded by the HGU, making it impractical to annotate such data. Conversely, the central aggregator possesses powerful computing resources and can effectively accomplish this work. Semi-supervised Learning (SSL) aims at leveraging unlabeled data to enhance model training by learning classification boundaries within these unlabeled samples and evaluating their proximity to labeled ones. Consequently, this approach strengthens both the robustness and generalization ability of models. The advantages of SSL primarily are in two aspects: (1) enhancing the robustness of TC classification models; (2) mitigating loss in model generalization caused by domain differences—for instance, traffic forwarded by different HGU devices may exhibit non-independent and identically distributed (non-IID) characteristics. In the subsequent section, we will propose an SSL-based method for traffic classification using FSSL.

\subsubsection{\textbf{Problem Formulation}}
\begin{enumerate}[label=\alph*)]
    \item Basic definitions: $D=[D_l, D_u]$; $D$ refers to the network traffic dataset, which consists of a labeled dataset $D_l$ and an unlabeled dataset $D_u$, $D_l \cap D_u = \phi$; $M$ and $N$ represent the total number of records in labeled and unlabeled datasets, respectively; $L=\{ l_1, l_2, \cdots, l_c \}$, refers to the dataset of the traffic classification label, where $l \in L, 0 \le c\le K$, and $K$ is the total number of application types; $F=F+F'$, is a collection of flow feature vectors; $F=[F_1, F_2, \cdots, F_M]$, refers to the set of labeled flow feature vectors; $F=[F_1', F_2', \cdots, F_N']$, refers to the set of unlabeled flow feature vectors; $F=F+F'=\{ f_l^i, f_u^j \}=\{ f_l^1, f_l^2, \cdots, f_l^n, f_u^1, f_u^2, \cdots, f_u^m, \}$, labeled/unlabeled flow feature vectors are denoted by $f_l^i$ and $f_u^j$, $0 \le i \le M$, $0 \le j \le N$.
    \item Definitions related to network traffic: 
    \begin{itemize}
        \item \textbf{flow:} A flow is identified by a five-tuple consisting of the source address, destination address, source port, destination port, and TCP/UDP protocol. $T=\{ t_1, t_2, \cdots, t_{m+n} \}$, represents a set of flows.
        \item \textbf{flow feature:} It includes packet-level features, flow-level features, and statistical features, formally defined as $f=\{ f^1, f^2, \cdots, f^{78} \}$. $f$ is the flow feature vector, composed of a total of 78 feature sub-items [13], which consist of the following three types of features:
        \begin{itemize}
            \item Packet-level features: The temporal and spatial features with packets as the granularity, including packet payload characteristics, packet length-related features, and time-related features. For example, packet length, inter-arrival time between packets, etc.
            \item Flow-level features: The temporal and spatial features of flows, with flows as the granularity, including flow length, flow duration, number of packets in a flow, and so on.
            \item Statistical features: The expectation, variance, maximum value, and minimum value of the relevant feature.
        \end{itemize}
\end{itemize}

    Table \ref{table_feature set} presents an example collection of traffic features for FedEdge AI-TC. From the example, it can be observed that specific feature entries involve a high level of user privacy.
    \begin{table*}[htbp]
    \centering
    \caption{A typical example (partial) of a network traffic feature set}
    \label{table_feature set}
    \begin{tabular}{llll}
        \hline \text { Flow Attributes } & \text { Definition } & \text { Catogory } & \text { Description } \\
        \hline 
        \text { Domain Name } & \text { DNS/SNI in TLS } & \text { Payload related } & \text { domain.com, which is applicable to applications such as HTTP/HTTPS. } \\
        \text { TCP slide\_win } & \text { TCP Slide Window } & \text { Packet related } & \text { TCP flow control parameters } \\
        \text { TLS\_handshake } & \text { TLS handshake packet information } & \text { Payload related } & \text { Handshake types, cipher suites, content types, key length, etc. } \\
        \text { Total Fwd Pkts } & \text { Packet length sequence } & \text { Packet related } & \text { The sequence of packet lengths in the flow. } \\
        \text { Pkt IAT Min } & \text { Packet arrival time } & \text { Packet related } & \text { The sequence of arrival times of packets in the flow. } \\
        \text { Flow Len } & \text { Flow length correlation } & \text { Flow related } & \text { The total number of bytes in the flow per unit of time. } \\
        \text { Flow Duration } & \text { Flow duration } & \text { Flow related } & \text { The duration of the TCP flow. } \\
        \hline
    \end{tabular}
    \end{table*}

    \begin{figure}[htbp]
    	\centerline{\includegraphics[width=9cm, height=4cm]{ 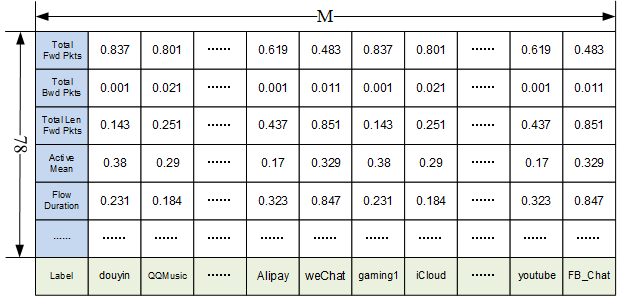}}
    	\caption{The Example of L\_FAM.}
    	\label{fig_FAM}
    \end{figure}

    \item Flow Features Matrix(FAM): 
    FAM comprises flow feature vectors and application category labels, formally defined as $FAM=[L\_FAM, U\_FAM]$. $L\_FAM$ refers to the labeled flow feature matrix, $L\_FAM=F+L=[[F_1, F_2, \cdots, F_M]^T,L^T]$. $U\_FAM$ refers to the unlabeled flow feature matrix, $U\_FAM=F+L=[F_1', F_2', \cdots, F_N']]$. Fig.~\ref{fig_FAM} represents an example of $L\_FAM$.

\end{enumerate}

\subsubsection{\textbf{VAE}}~{}

As we all know, an Autoencoder (AE) is specifically designed to acquire a low-dimensional latent representation of samples by constructing an encoder and a decoder. It is commonly employed for tasks such as data compression or generation. However, due to its sole focus on learning the encoding of the sample itself, AE-based models usually show weak generalization. Therefore, AE's capacity to capture the underlying data distribution needs to be improved. In addition, network traffic consistently exhibits characteristics such as large scale, dynamic, and heterogeneity. A conventional AE model usually fails to fully reconstruct comprehensive network traffic even when provided with extensive datasets. Consequently, accurate classification becomes imperative when encountering traffic samples beyond the dataset.

\begin{figure}[htbp]
	\centerline{\includegraphics[width=9cm, height=4cm]{ 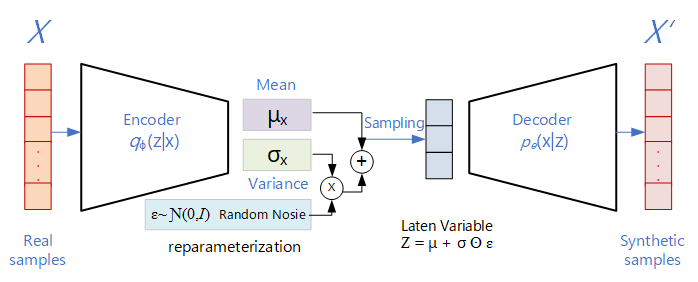}}
	\caption{The General Architecture of VAE.}
	\label{fig_VAE}
\end{figure}

Variational Autoencoder (VAE) is an extension of Autoencoder (AE). More precisely, it is a generative model widely employed for unsupervised pre-training of unlabeled data. Its excellent ability to learn the latent distribution enables the model to acquire strong generalization capabilities during subsequent fine-tuning. As depicted in Fig.~\ref{fig_VAE}, the core concept behind VAE lies in learning the implicit representation of actual samples $X$ and the implicit distribution from these samples to generated samples $X'$. This approach enhances the robustness of the model in learning implicit feature representations. It estimates the overall data distribution by constructing a generative model $p_\varphi(Z|X)$ based on data samples. However, commonly used methods for estimating data distributions rely on maximum likelihood estimation through parameter estimation techniques. Assuming that the distribution of data samples $P(X)$ follows a Gaussian distribution $N(\mu, \sigma^2)$, this transforms the statistical problem of generating models into a parameter estimation problem. The remaining challenge lies in fitting the distributions of encoder $q_\phi(Z|X)$ and decoder $p_\varphi(Z|X)$. While autoencoders (AE) typically learn data distributions by minimizing reconstruction loss like Mean Square Error (MSE), this type of fitting primarily occurs at a sample level within datasets. It fails to capture underlying data distributions effectively. In contrast, VAE employs KL divergence as a measure for quantifying differences between two distributions, also known as relative entropy. Therefore, we can define our loss function as Eq.~\ref{eq_Loss}.

\begin{equation}
    \label{eq_Loss}
    Loss = L(X, X') + \sum_j KL(q_j(Z|X) \parallel p(Z))
\end{equation}

The loss function consists of two components: the first component is the reconstruction loss, and the second component is the KL divergence between the proper distribution and the distribution we have chosen. VAE aims to minimize this relative entropy, as expressed in Eq.~\ref{eq_L(x)1}.
\begin{equation}
    \label{eq_L(x)1}
    \begin{aligned}
        L(x) & = E_{z~q(z|x)} \log \frac{p(z, x)}{q(z|x)} \\
             & = \log p(x) - KL(q(z|x) \parallel p(z|x)) \\
    \end{aligned}
\end{equation}
$L(x)$ is called the Variational Lower Bound. We aim to optimize this lower bound, as the closer it is to $\log p(x)$, the smaller the KL divergence. In this case, $q_\phi (X|Z)$ can estimate $p_\varphi (Z|X)$ more accurately.

The VAE further decomposes the sampling of $z$ into two parts: one consists of fixed values such as the standard deviation $\sigma$ and the mean $\mu$, and the other is a random Gaussian noise $\epsilon$ . After applying the reparameterization trick, we can rewrite $L(x)$ as Eq.~\ref{eq_L(x)2}:
\begin{equation}
    \label{eq_L(x)2}
    L(x) = \frac{1}{2} \sum_{i=1}^J (1 + \log (\sigma_j^2) - \mu_j^2) + \frac{1}{L} \sum_{i=1}^L \log p(x|z_i)
\end{equation}

The optimization of the variational lower bound of $L(x)$ implies that, while ensuring that the $Z$ values generated by the encoder conform to a prior Gaussian distribution, the decoder can maximize the possibility of reconstructing the original $X$.

\subsubsection{\textbf{VAE based Unsupervised Learning Algorithm for Network TC}}~{}\label{sec:un-VAE}

As shown in Alg.\ref{alg:vae}, the entire algorithm mainly consists of the following three steps: 
\begin{enumerate}[label=\alph*)]
    \item \textbf{Define the hyperparameters:} Input dimension is $input\_dim$; Hidden layer dimensions are $hi\_dim$ for $0<i<L$, where $L$ is the number of hidden layers; Dependent variable dimension is $z\_dim$; Batch size is $batch\_size$; Number of epochs for training is $num\_epochs$.
    \item \textbf{Dataset:} $X \in D_u$, $X$ is equivalent to a $U\_FAM$ with $input\_dim \times batch\_size$.
    \item \textbf{Model construction and training:} 
    \begin{itemize}
        \item \emph{Define the exact architectures of Encoder and Decoder.} This includes determining the number of layers, $input\_dim$, $hi\_dim$, $z\_dim$, and the loss function. The Encoder maps $X$ into the latent space $Z$, while the Decoder maps the randomly sampled $z$ from $Z$ into the data space $X'$. The ultimate goal is to make $X$ and $X'$ as close as possible.
        \item \emph{The feed-forward propagation process from Encoder to Decoder.} 
        \begin{itemize}
            \item In this context, the input data $X$, referred to as $U\_FAM$, is fed into the Encoder. After sequential computations, the mean and variance of the posterior distribution $Z$ in the latent space are obtained.
            \item The technique of reparameterization is used to sample a latent variable $z$ from $Z$, i.e., $Z = \mu + \sigma \odot \varepsilon$, where $\varepsilon \sim N(0, 1)$, and it is then fed into the Decoder.
            \item The decoder performs layer-by-layer computations to obtain the reconstructed output $X'$ of the input data $X$, which can be expressed as $U\_FAM'$.
            \item Calculate the reconstruction error and KL divergence based on Eq.~\ref{eq_Loss} to obtain.
        \end{itemize}
        \item \emph{Backpropagation and Optimization.} By iterating through $num\_epochs$ and utilizing the optimizer defined, the VAE model is trained in a loop to optimize the model parameters with the goal of minimizing $L(x)$.
    \end{itemize}
\end{enumerate}

\begin{figure*}[ht]
	\centering\includegraphics[scale=0.6]{ 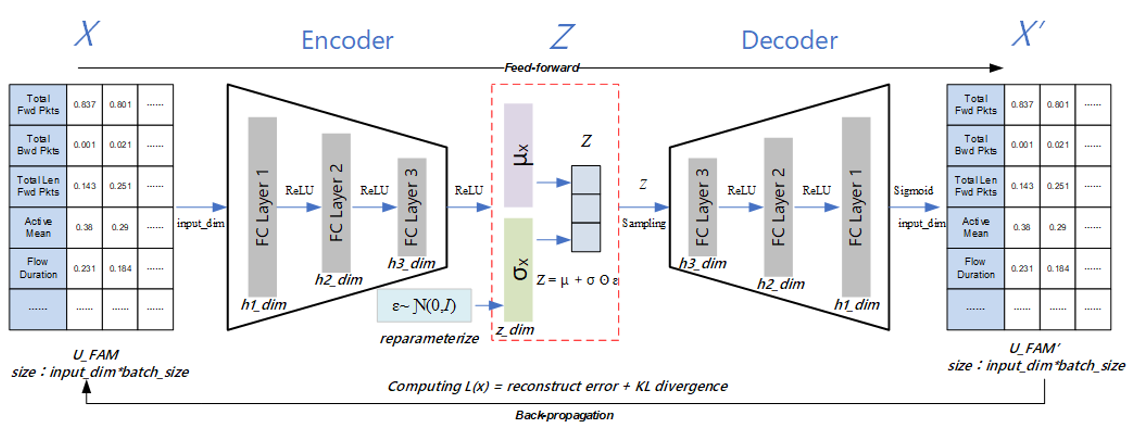}
	\caption{The VAE Model Architecture of FedEdge AI-TC.}
	\label{fig_VAE_Model} % add a label for reference
\end{figure*}

\begin{algorithm}[htbp]
\caption{VAE based Unsupervised Learning Algorithm for Network Traffic Classification}\label{alg:vae}
\renewcommand{\algorithmicrequire}{\textbf{Input:}}
\renewcommand{\algorithmicensure}{\textbf{Output:}}
\begin{algorithmic}[1]
\REQUIRE ~{}
\\Training data: $U\_FAM$, $X \in D_u$; 
\\Training parameters: $imput\_dim$, $h_i\_dim$, $0<i<L$, $z\_dim$, $batch\_size$, $num\_epochs$, $learning\_rate$, $optimizer$;
\ENSURE VAE based unsupervised model.
\STATE Initial training dataset $X \in D_u$;
\STATE Define the exact architectures of Encoder and Decoder;
\FOR{epoch in $num\_epochs$}
    \FOR{epoch batch of $U\_FAM$ of $X$}
    \STATE For each training samples $X \in D_u$:
    \STATE Encoder: computing the mean-$\mu_x$ and variance-$\sigma_x$ of $Z$;
    \STATE Encoder: using reparameterize skill to sample $z$ from $Z$; $z = \mu_x + \sigma_x \odot \varepsilon$; $\varepsilon$ from $N(0, 1)$;
    \STATE Decoder:computing by layer-wise to get the reconstruct output $X'$ of $X$, which is $U\_FAM'$;
    \STATE Compute the loss of reconstruct and KL-divergence using equation(1);
    Update the weights and bias;
    \ENDFOR
\ENDFOR
\end{algorithmic}
\end{algorithm}

The structure of the unsupervised model for network traffic based on VAE is shown in Fig.~\ref{fig_VAE_Model}, and detailed parameters are provided in Table \ref{table_detailed parameters}.

\begin{table}[htbp]
\centering
\caption{Detailed parameters of FedEdge AI-TC}
\label{table_detailed parameters}
\begin{tabular}{m{3.5cm} m{1cm} m{3cm}}
    \hline Parameter name & Parameter value & Parameter Interpretation \\
    \hline 
    input\_dim & 78 & Model input dimension \\
    Layers of Encoder/Decoder & 3 & Number of layers in the encoder and decoder \\
    $h_1\_dim$, $h_2\_dim$, $h_3\_dim$ & 78,64,32 & Dimension of each layer\\
    loss function for Encoder & ReLU & Loss function \\
    loss function for Decoder & ReLU & Loss function \\
    loss function for Decoder's output & Sigmoid & Loss function \\
    batch\_size & 128 & Batch size \\
    learning rate & 0.01 & Learning rate \\
    \hline
\end{tabular}
\end{table}

\subsubsection{\textbf{VAE+CNN based Semi-supervised Learning Algorithm for Network TC}}~{}

\begin{figure}[htbp]
	\centerline{\includegraphics[width=8cm, height=4cm]{ 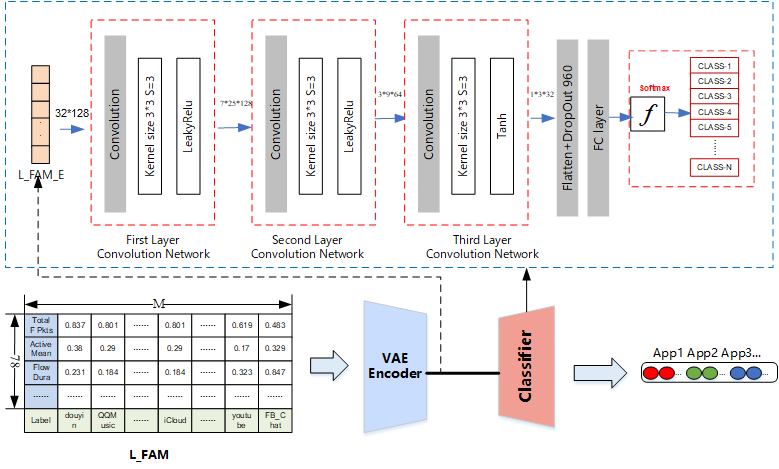}}
	\caption{The VAE+CNN Semi-supervised Model Architecture of FedEdge AI-TC.}
	\label{fig_VAE_CNN_Model}
\end{figure}

As illustrated in Fig.\ref{fig_VAE_CNN_Model}, there are three parts in the semi-supervised based network traffic classifier: the encoder of VAE model, CNN, and softmax classifier. The labeled data, i.e., L\_FAM is fed into the Encoder of the VAE model obtained from Section.\ref{sec:un-VAE}, then the output of the VAE encoder is subsequently fed into a three layers CNN model with a softmax classifier, which is concatenated with the VAE encoder. Finally, the decision results will be outputted for classification. The overall process is commonly referred to as \textbf{\emph{Fine-Tuning}}. Due to the limited space of this paper, we do not provide the detail of the CNN classifier, which can be found in our previous work\cite{45}.

\subsection{The Model Compression Method Based on Interpretation}

\subsubsection{Model Interpretation}
We propose to implement an interpretable framework for deep learning traffic classification models based on SHAP values, which are mainly used to quantify the contribution of each feature to the model prediction. The basic design idea is to calculate the marginal contribution SHAP value when features are added to the model so that the importance of the features can be interpreted according to the SHAP value, which is calculated as Eq.~\ref{eq_f(xij)} in this paper. Suppose the i-th sample of sample set M is $x_i$, the j-th feature of sample $x_i$ is $x_{ij}$, $f (x_{ij})$ is the shapely value of $x_{ij}$.
\begin{equation}
    \label{eq_f(xij)}
    f(x_{ij})=\sum_S \frac{\left | S \right |!(p-\left | S \right |-1)!}{p!} (v_x(S\cup \{x_j\})-v_x(S))
\end{equation}
where $S \subseteq \{x_1,\cdots,x_p\} \setminus \{x_j\}$, $\{x_1,\cdots,x_p\}$ is the set of all possible input features excluding $\{x_j\}$, $p$ is the number of features of the sample, and $v_x(S)$ is the prediction result of the feature subset $S$.

\begin{figure}[htbp]
	\centerline{\includegraphics[scale=0.3]{ 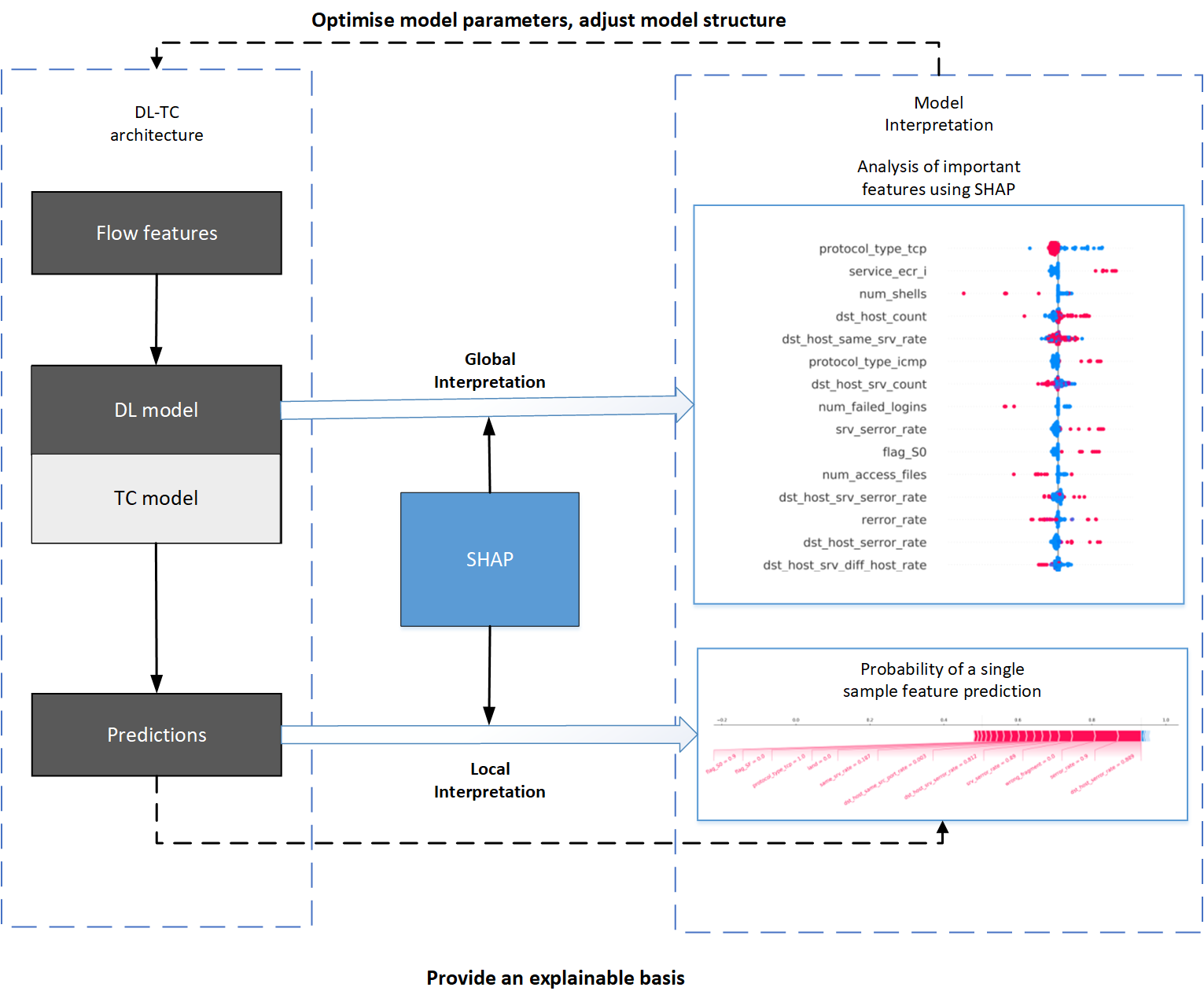}}
	\caption{The Architecture of Model Interpretation.}
	\label{fig_interpret}
\end{figure}

The architecture of Model Interpretation is shown in Fig.~\ref{fig_interpret}. The left part is the traditional structure of the traffic classification model, and the process shown on the right allows for the interpretability of the traffic classification model and the optimization of the structure and parameters of the traffic classification model. This framework is divided into a local interpretation and a global interpretation.

\textbf{\emph{Local interpretation}} means that for each data instance, the contribution of each feature to its predicted outcome is calculated and presented visually. The formula for calculating the local interpretation is as follows: 
\begin{equation}
    \label{eq_yi}
    y_i=y_{base}+f(x_{i1}+f(x_{i2}+\cdots+f(x_{ij})
\end{equation}
where $y_i$ is the predicted value of the model for sample $x_i$, $y_{base}$ is the mean of all sample evaluated values.

As for \textbf{\emph{Global interpretation}}, firstly, a matrix of feature SHAP values is calculated, with one instance per row and one feature per column. Secondly, in the traditional global interpretation, the feature j's contribution is obtained by summing the shapely mean of feature j for all samples with Eq.~\ref{eq_f(xj)}. And then, SHAP values are sorted in descending order to obtain the importance of the model features. 
\begin{equation}
    \label{eq_f(xj)}
    f(x_j)=\sum_{i=1}^M f(x_{ij})
\end{equation}

% However, using the SHAP average to rank the importance of features is inaccurate in cases where multiple features jointly influence the classification results. Therefore, in order to solve the problem of averaging feature contributions in the SHAP calculation, this paper uses the Eq.~\ref{eq_chisqure-test} to calculate the correlation between features and results by means of a chi-square test, and adds the normalized chi-square test value as a weighting factor. 
% \begin{equation}
%     \label{eq_chisqure-test}
%     x^2=\sum \frac{(A-T)^2}{T}
% \end{equation}
% where $A$ is the actual frequency, $T$ is the theoretical frequency, and $x^2$ is the chi-square test value.
    
% The improved formula for calculating the global SHAP value in this paper is as Eq.~\ref{eq_chif(xj)}.
% \begin{equation}
%     \label{eq_chif(xj)}
%     \begin{aligned}
%         & f(x_j)=\sum_{i=1}^M \lambda f(x_{ij})\\
%         & \lambda=\frac{x^2-x_{min}^2}{x_{max}^2-x_{min}^2}
%     \end{aligned}
% \end{equation}

% The SHAP method, improved using the chi-square test, can increase the weight of sample classification features when multiple feature factors are combined. It is able to make full use of the classification features learned by the model, thus validating the reliability of the model.

\subsubsection{Model Pruning}
To address the issue of how to compress models to make them suitable for training and inference on resource-limited devices, in particular, we will focus on pruning, an easier-to-implement model compression technique. Model pruning is based on an underlying assumption: 'weight contribution,' which means that not every weight contributes equally to the output prediction. 

Therefore, the basic design idea of model pruning in this paper is to rank the feature importance by global interpretation, after which the importance ranking of the convolutional kernels is calculated using the causal evaluation mechanism. The convolutional kernels with importance below a threshold are filtered out and pruned.

\section{Experimental Evaluation and Discussion}\label{sec:evaluation}

\subsection{Evaluation Settings and Chosen Datasets}

We conducted the experimental evaluation on two datasets. One is a benchmarking public dataset, ISCX-VPN2016. The other is a private dataset built by ourselves. The latter comprises six popular applications and background traffic from terminals collected in the campus network scenario, including Bilibili, QQ music, Honor of Kings, Teamfight Tactics, and Game for peace. These apps cover the five popular app categories of video, music, Moba games, First Person Shooting (FPS) and Role-playing game (RPG). To collect the data, we used a semi-automatic web traffic generation tool. We leveraged an automated traffic generator to collect traffic from Bilibili and QQ music, with 13,314 flows from Bilibili and 20703 flows from QQ music. For the other interactive games, we chose to collect them manually. We used PCAPDroid to mark the network traffic as it occurred at the endpoint and also at the router, where the two were compared and filtered. The network flow features were calculated by CICFlowMeter for each application's PCAP files. In addition, this experiment also provides the statistics of the background traffic information, which contains information about network location services, security components, and syslog services. Table \ref{Datasets} shows the exact number of flows.

\begin{table}[htbp]
\caption{The private experimental dataset}
\label{Datasets}
\centering
\begin{tabular}{ccc}
\hline
Applications Name           & Type  & Number \\ \hline
bilibili       & Video & 13314  \\ 
QQ music       & Music & 20703  \\ 
Honor of Kings  & Moba  & 9475   \\ 
Teamfight Tactics & RPG   & 14005  \\ 
Game for peace & FPS   & 7763   \\ 
Background     & Log   & 13017  \\ \hline
\end{tabular}
\end{table}

The experimental environment is AMD Ryzen 3600, 16GB RAM, NVIDIA GTX 1660, CUDA 7.5, CDNN10.5. In this paper, Python3 is the primary programming language. The following is a description of the evaluation metrics: Precision, Recall, F1, Accuracy, and AUC.

Time-complexity-related metrics about training like training time are not included in this paper because we think those are highly dependent on the hardware resources.

\subsection{Performance Evaluation}
The training process of the VAE part in the VAE+CNN (E-CNN) model is divided into two parts. Firstly, the labels were removed from the datasets, which can be acted as the unlabeled flows for the unsupervised learning training. After training, one can save the trained encoder in the VAE model for further semi-supervised learning. The training process for the CNN part mainly aim to convert the labeled data into digital encoding for fine-tuning. 

In the single CNN model, we process the dataset and divide it into training and testing sets according to ratio, which is then trained and evaluated in the CNN.

\begin{figure}[htbp]
	\centerline{\includegraphics[scale=0.6]{ 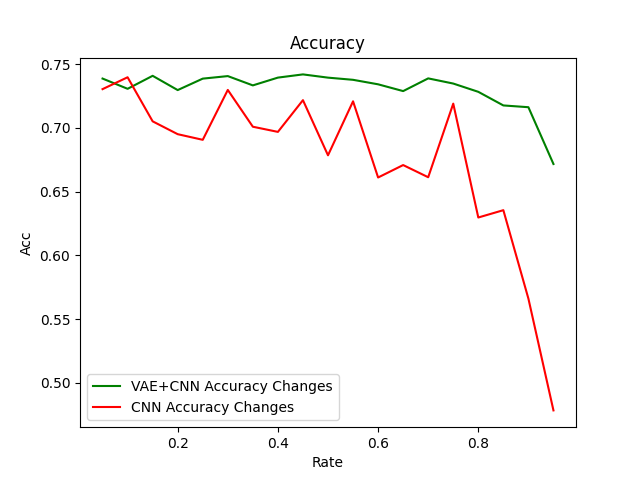}}
	\caption{Accuracy of different rates in real-life scenarios.}
	\label{fig_CNN_Accuracy}
\end{figure}

\begin{figure}[htbp]
	\centerline{\includegraphics[scale=0.6]{ 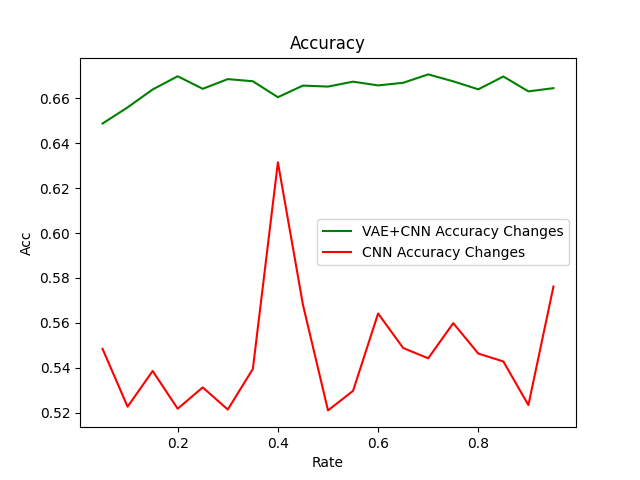}}
	\caption{Accuracy of different rates in public dataset.}
	\label{fig_Accuracy}
\end{figure}

For the E-CNN and single CNN models, we obtained the following two diagrams by adjusting the ratio of the dataset to the training and testing sets. The horizontal axis in the diagram represents the partitioning ratio. For example, 0.2 means that 80$\%$ of the entire dataset is allocated to the training set, and the remaining 20$\%$ is allocated to the testing set. The vertical axis represents the accuracy of the model training results under this partition. Fig.~\ref{fig_CNN_Accuracy} displays the model results of E-CNN and single CNN in real-life scenarios, while Fig.~\ref{fig_Accuracy} shows the results in the public dataset.

% \begin{figure}[h]
%     \subfigure[Real-life scenario.]{
%     \includegraphics[width=.45\columnwidth]{ acc(2).png}
%     \label{fig_CNN_Accuracy}
%     }
%     \subfigure[Public dataset.]{
%     \includegraphics[width=.45\columnwidth]{ fig12-acc.png}
%     \label{fig_Accuracy}
%     }
%     \caption{Accuracy of different rates}
% \end{figure}

According to the figures, as the partition ratio increases, meaning that the training set data decreases, the accuracy of both models tends to decrease and becomes similar, both at around 0.7. Within the partition ratio interval [0.2, 0.8], E-CNN achieves good results with higher accuracy than a single CNN.

\begin{figure}[htbp]
	\centerline{\includegraphics[scale=0.25]{ 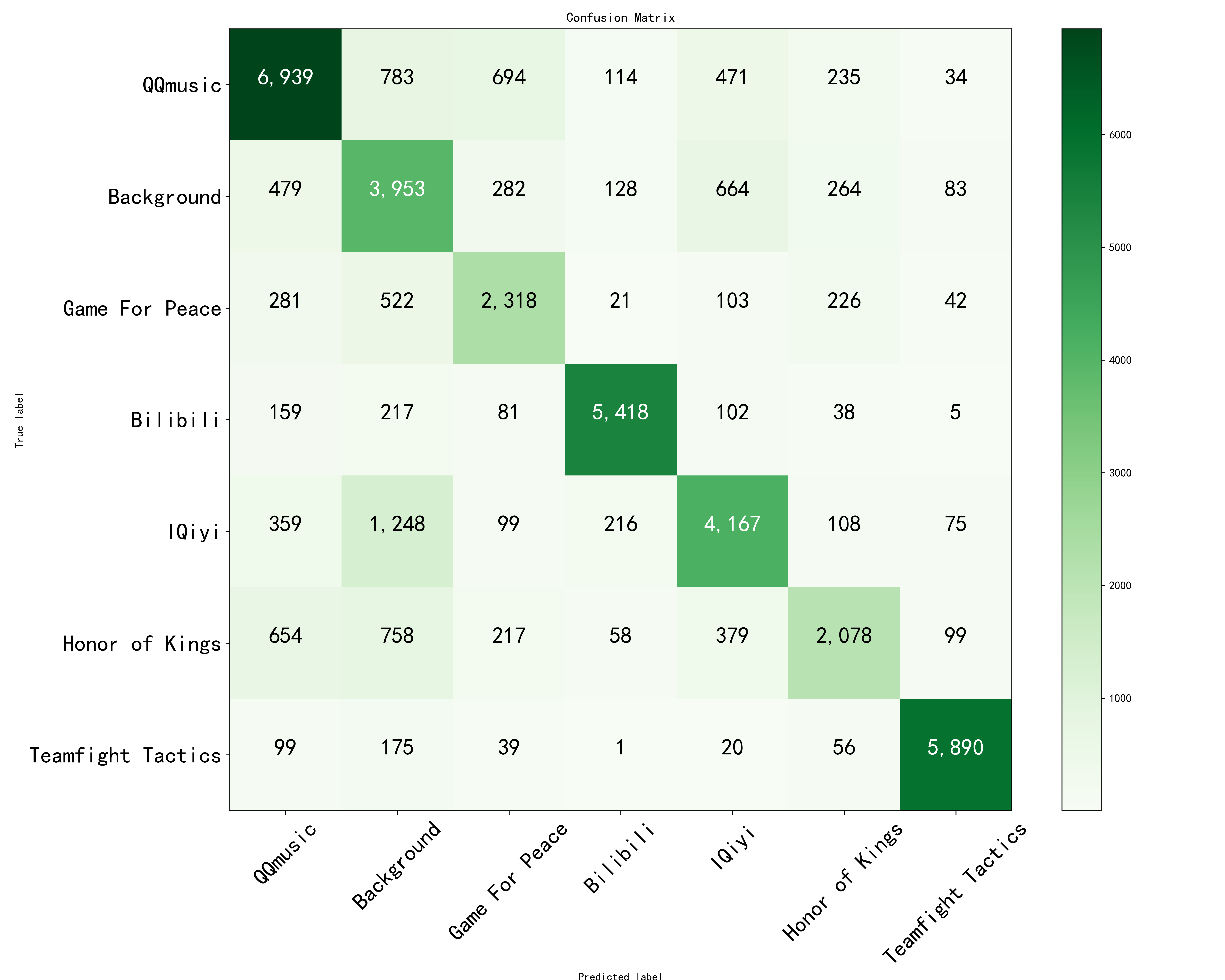}}
	\caption{VAE+CNN Confusion Matrix}
	\label{fig_VAE+CNN_confusionmatrix}
\end{figure}

\begin{figure}[htbp]
	\centerline{\includegraphics[scale=0.25]{ 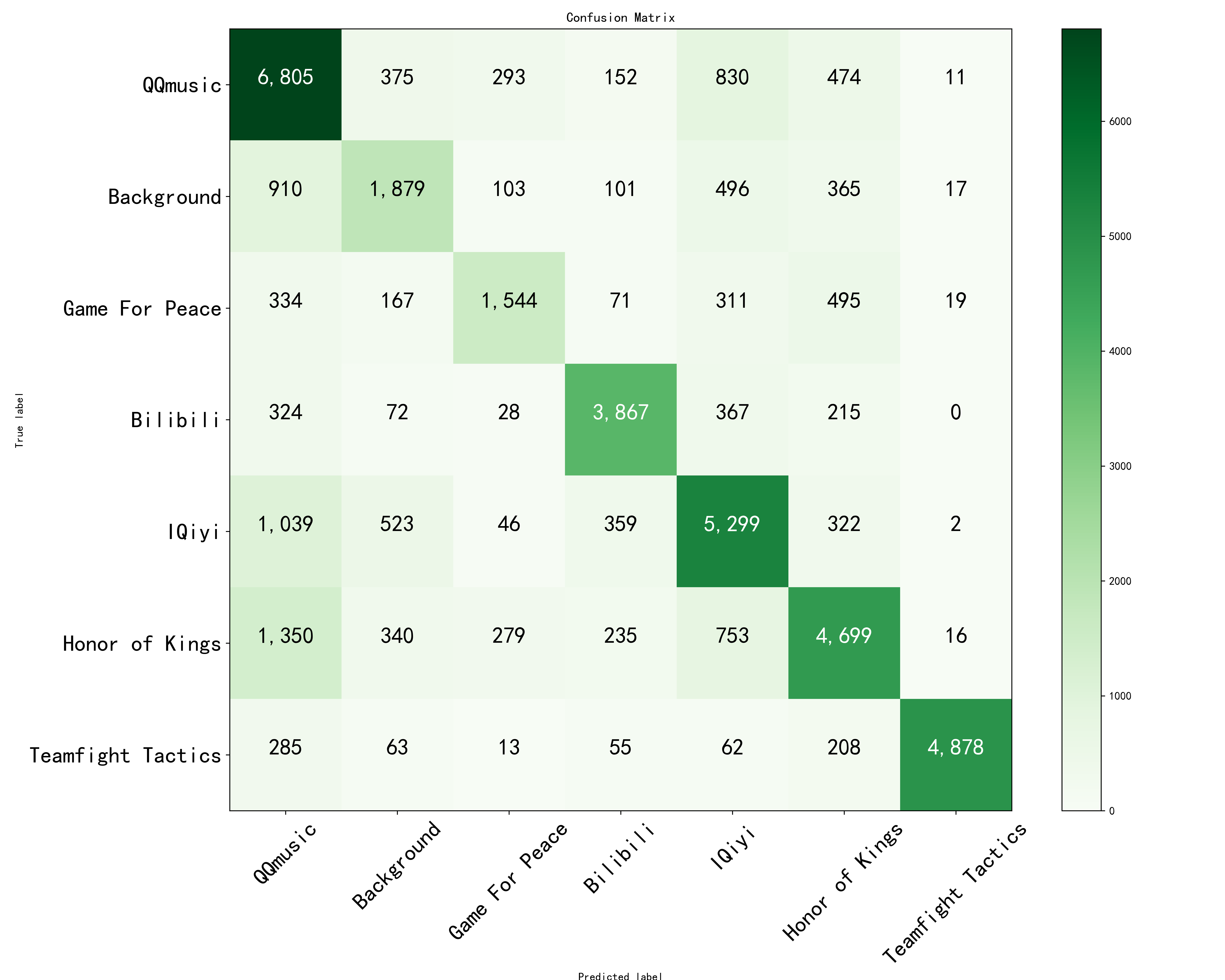}}
	\caption{CNN Confusion Matrix}
	\label{fig_CNN_confusionmatrix}
\end{figure}

We have set the partition ratio to 0.45. After training and testing, we obtained the respective Confusion Matrix and Classification Report for E-CNN and single CNN using this ratio. The Confusion matrix is shown in Fig.~\ref{fig_VAE+CNN_confusionmatrix} and Fig.~\ref{fig_CNN_confusionmatrix}. 
In the figures, the x-axis represents the prediction labels, the y-axis represents the actual labels, and the color intensity indicates the count of correct and incorrect predictions. According to the figure, under the partition with this ratio, E-CNN shows relatively accurate predictions for 'QQmusic,' 'Teamfight Tactics,' and 'Bilibili.' On the other hand, compared to CNN, E-CNN's predictions for 'QQmusic,' 'IQiyi,' and 'Teamfight Tactics' are relatively accurate, but there are also several errors.

% \begin{figure}[h]
%     \subfigure[VAE+CNN]{
%     \includegraphics[width=.45\columnwidth]{ fig9-VAE+CNN confusionmatrix.png}
%     \label{fig_VAE+CNN_confusionmatrix}
%     }
%     \subfigure[CNN]{
%     \includegraphics[width=.45\columnwidth]{ fig10-CNN-confusionmatrix.png}
%     \label{fig_CNN_confusionmatrix}
%     }
%     \caption{Confusion Matrix}
% \end{figure}

The classification report is presented in Table \ref{VAE+CNN-Classification-Report} and Table \ref{CNN-Classification-Report}. It is evident from the tables that, under this partition ratio, E-CNN demonstrates relatively accurate predictions for 'Teamfight Tactics' and 'Bilibili,' with prediction accuracy rates above 0.9. The prediction accuracy for other applications is also around 0.7, resulting in a total classification accuracy of 0.7422. In contrast, CNN's performance is notably inferior to that of E-CNN. The lowest prediction accuracy is approximately 0.5, and the highest prediction accuracy is only around 0.8. Its overall classification accuracy is merely 0.6989.

\begin{table}[htbp]
	\small
	\caption{Classification Report of VAE+CNN}
	\label{VAE+CNN-Classification-Report}
	\centering
\begin{tabular}{lllll}
\hline
                  & Precision & Recall & F1-score & Support \\ \hline
QQmusic           & 0.7736    & 0.7485 & 0.7609   & 9270    \\ 
Background        & 0.5163    & 0.6754 & 0.5852   & 5853    \\ 
Game For Peace    & 0.6214    & 0.6598 & 0.6401   & 3513    \\ 
Bilibili          & 0.9097    & 0.9000 & 0.9048   & 6020    \\ 
IQiyi             & 0.7056    & 0.6644 & 0.6843   & 6272    \\ 
Honor of Kings    & 0.6915    & 0.4897 & 0.5734   & 4243    \\ 
Teamfight Tactics & 0.9457    & 0.9379 & 0.9418   & 6280    \\ 
accuracy          &           &        & 0.7422   & 41451   \\ 
macro avg         & 0.7377    & 0.7251 & 0.7272   & 41451   \\ 
weighted avg      & 0.7515    & 0.7422 & 0.7434   & 41451   \\ \hline
\end{tabular}
\end{table}

\begin{table}[htbp]
        \small
	\caption{Classification Report of CNN}
	\label{CNN-Classification-Report}
	\centering
\begin{tabular}{lllll}
\hline
                  & Precision & Recall & F1-score & Support \\ \hline
QQmusic           & 0.6160    & 0.7612 & 0.6809   & 8940    \\ 
Background        & 0.5496    & 0.4854 & 0.5155   & 3871    \\ 
Game For Peace    & 0.6696    & 0.5250 & 0.5885   & 2941    \\ 
Bilibili          & 0.7990    & 0.7936 & 0.7963   & 4873    \\ 
IQiyi             & 0.6527    & 0.6982 & 0.6747   & 7590    \\ 
Honor of Kings    & 0.6933    & 0.6125 & 0.6504   & 7672    \\ 
Teamfight Tactics & 0.9869    & 0.8767 & 0.9285   & 5564    \\ 
accuracy          &           &        & 0.6989   & 41451   \\ 
macro avg         & 0.7096    & 0.6789 & 0.6907   & 41451   \\ 
weighted avg      & 0.7059    & 0.6989 & 0.6989   & 41451   \\ \hline
\end{tabular}
\end{table}

Based on the experimental results presented above, we can conclude that when the ratio of labeled data to total data falls within the range of [0.2, 0.8], the VAE+CNN method proposed in this paper can be employed for classifying and recognizing network application traffic. At this range, the recognition outcomes outperform those achieved by solely training with CNN, and the classification accuracy is commendable.

\begin{figure}[htbp]
    \centerline{\includegraphics[scale=0.3]{ 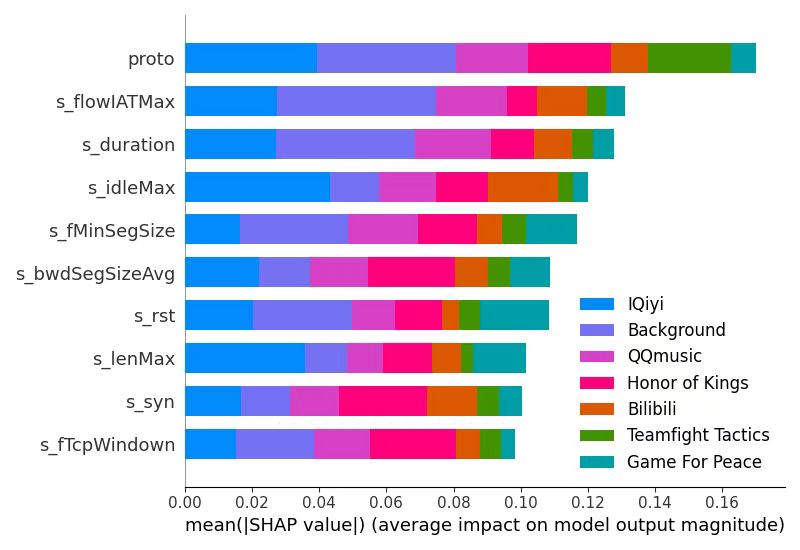}}
    \caption{Summaryplot of Global Interpretation.}
    \label{fig_summaryplot}
\end{figure}

Fig.~\ref{fig_summaryplot} visualizes the importance of ten features for the classification results. These ten features are the most significant ones contributing to the classification outcomes. For instance, in the case of the applications Bilibili and IQiyi, the most crucial feature for the classification result is s\_idleMax, representing the maximum idle time of any neighboring packet in a stream. Notably, video streams exhibit distinct characteristics on this feature due to their continuous packet sending and shorter inter-packet idle time, which differentiates them from game streams.

\begin{table}[htbp]
    \caption{Baseline Model vs. XAI-Pruning Model}
    \label{table_pruned}  
    \centering
    \begin{tabular}{ccc}
    \hline
     & Baseline model & XAI-Pruning model \\
    \hline
    Size of zipped file & 634163 bytes & 207292 bytes \\
    Accuracy of prediction & 0.75 & 0.69 \\
    Inference Time & 2.78s & 1.12s \\
    \hline
    \end{tabular}
\end{table}

The XAI-Pruning model effectively filters out unimportant weights and parameters, substantially reducing the number of parameters and the overall model size. As a result, the model becomes more compact, enabling significantly faster inference and quicker predictions. On the other hand, the baseline model, which includes all weights and parameters, generally exhibits higher prediction accuracy due to its capacity to capture more intricate model details and complexity.

As depicted in Table \ref{table_pruned}, the baseline model size is approximately three times larger than the XAI-Pruning model, and the inference time is doubled. However, despite these changes, both models still exhibit comparable prediction accuracy without significant differences. Consequently, the pruning method proposed in this paper effectively reduces the model's compression size and inference time to some extent while having a relatively minor impact on prediction accuracy.

\section{Conclusion and Future work}\label{sec:conclusion}
This paper presents the FedEdge AI-TC approach for trusted Federated Learning (FL) based efficient Network TC in 5G CPE/HGU. Firstly, FedEdge AI-TC effectively protects the data privacy of network subscribers by proposing an FL-based framework of local training, model parameters iterating, and centralized training. Secondly, a semi-supervised TC algorithm based on Variational Auto-Encoder (VAE) and convolutional neural network (CNN) is designed to reduce data dependence while keeping the TC accuracy. Finally, XAI-Pruning, an AI model compression method, combined with the DL model interpretability, is proposed to condense the model and interpret it globally and locally to achieve light-weighted AI-TC model deployment while building the trust in their decision of network operators. To demonstrate the efficiency of the proposed method, we conducted some experimental evaluations on commonly used public benchmark datasets and real network datasets. The results show that FedEdge AI-TC can outperform the benchmarking methods regarding the accuracy and achieve excellent TC performance of model inference on 5G CPE/HGU with limited computing resources, which effectively protects the users' privacy and improve the model's credibility.

However, besides reliability, robustness, and generalization are still two important topics when handling network traffic classification, especially using ML/DL. In the future, we will continuously focus on how to leverage ML/DL algorithms like generative models or large language models to enhance the reliability, robustness, and generalization of network traffic classification models.

\section*{Acknowledgment}
\noindent The paper is supported by National Natural Science Fundation (General Program) of China under Grant 61972211

\ifCLASSOPTIONcaptionsoff
  \newpage
\fi

% trigger a \newpage just before the given reference
% number - used to balance the columns on the last page
% adjust value as needed - may need to be readjusted if
% the document is modified later
%\IEEEtriggeratref{8}
% The "triggered" command can be changed if desired:
%\IEEEtriggercmd{\enlargethispage{-5in}}

% references section
\renewcommand\refname{Reference}
\bibliographystyle{IEEEtran}
\bibliography{IEEEfull,Reference}

% can use a bibliography generated by BibTeX as a .bbl file
% BibTeX documentation can be easily obtained at:
% http://mirror.ctan.org/biblio/bibtex/contrib/doc/
% The IEEEtran BibTeX style support page is at:
% http://www.michaelshell.org/tex/ieeetran/bibtex/
%\bibliographystyle{IEEEtran}
% argument is your BibTeX string definitions and bibliography database(s)
%\bibliography{IEEEabrv,../bib/paper}
%
% <OR> manually copy in the resultant .bbl file
% set second argument of \begin to the number of references
% (used to reserve space for the reference number labels box)

% biography section
% 
% If you have an EPS/PDF photo (graphicx package needed) extra braces are
% needed around the contents of the optional argument to biography to prevent
% the LaTeX parser from getting confused when it sees the complicated
% \includegraphics command within an optional argument. (You could create
% your own custom macro containing the \includegraphics command to make things
% simpler here.)
%\begin{IEEEbiography}[{\includegraphics[width=1in,height=1.25in,clip,keepaspectratio]{mshell}}]{Michael Shell}
% or if you just want to reserve a space for a photo:

\begin{IEEEbiography}[{\includegraphics[height=1.25in,clip,keepaspectratio]{ 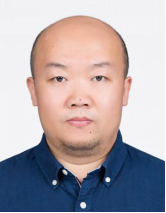}}]{Pan Wang} (M'18) received the BS/MS/Ph.D. degree in Electrical \& Computer Engineering from Nanjing University of Posts\&Telecommunications, Nanjing, China, in 2001, 2004, and 2013, respectively. He is currently a Full Professor at Nanjing University of Posts \& Telecommunications, 
Nanjing, China. His research interests include AI-powered networking and security in B5G/6G/IoT/Smart Grid/CFN, and AI-enabled big data analysis. From 2017 to 2018, he has been a visiting scholar at University of Dayton (UD) in the Department of Electrical and Computer Engineering, OH, USA. He served as a TPC member of IEEE CyberSciTech Congress. He is also a reviewer for several
journals, including IEEE Transaction on Network and Service Management, IEEE Transaction on  EMERGING TOPICS IN COMPUTATIONAL INTELLIGENCE, IEEE Internet of Things Journal, IEEE Journal on Selected Areas in Communications,
IEEE ACCESS, Computer Networks, Computer\&Security, Computer Communications, Engineering Applications of Artificial Intelligence, Big Data Research, etc. 
\end{IEEEbiography}

\begin{IEEEbiography}
[{\includegraphics[height=1.25in,clip,keepaspectratio]{ 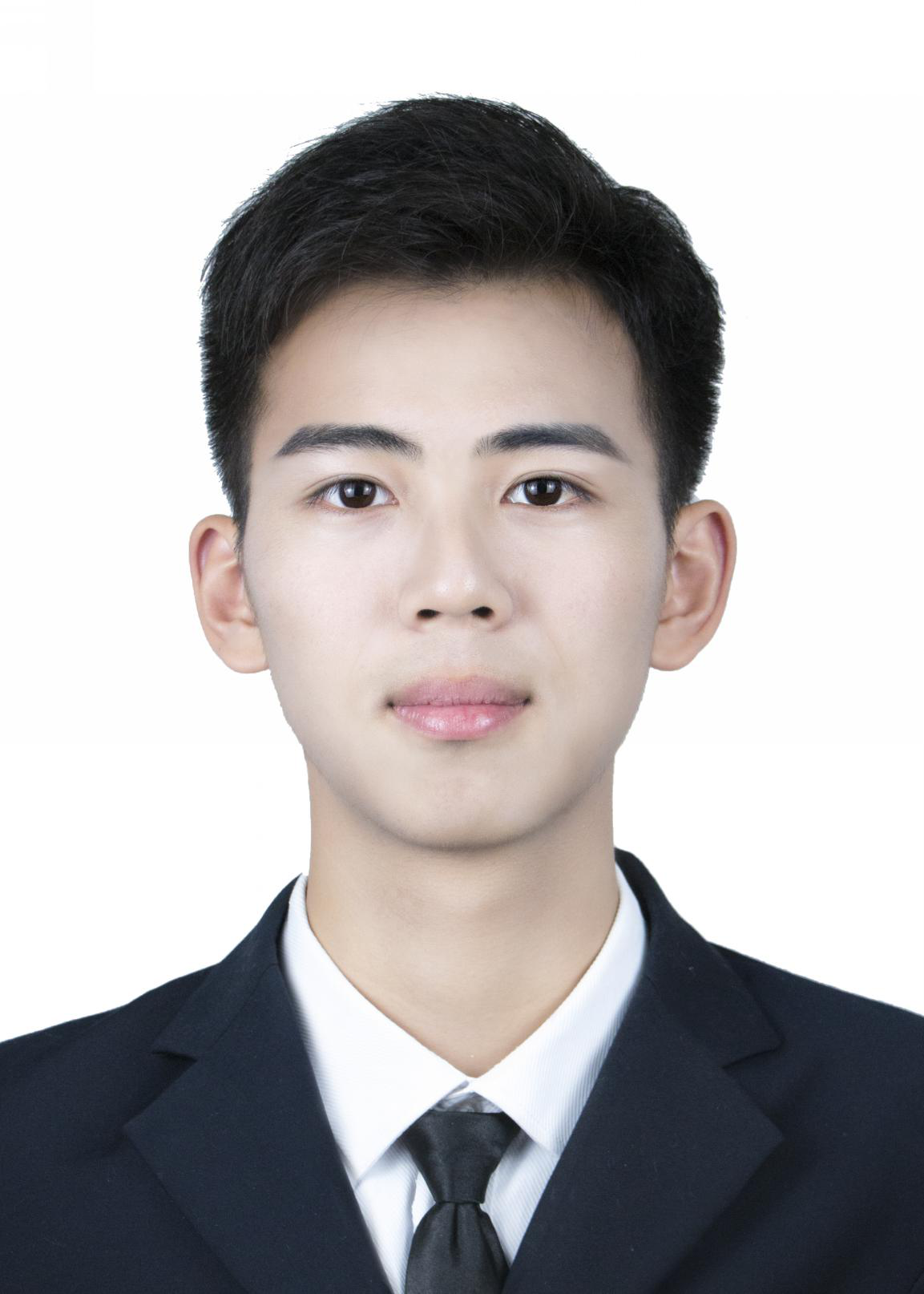}}]{Zeyi Li} is currently pursuing the Ph.D. degree in Cyberspace Security at Nanjing University of Posts and Telecommunications. He was born in Soochow, Jiangsu, China, in 1997. He received his bachelor's degree in mathematics in 2019 and received M.S. degree in computer science in 2022. His research interests include network security, communication network security, anomaly detection and analysis, deep packet inspection, and graph neural networks.
\end{IEEEbiography}
\begin{IEEEbiography}
[{\includegraphics[height=1.25in,clip,keepaspectratio]{ 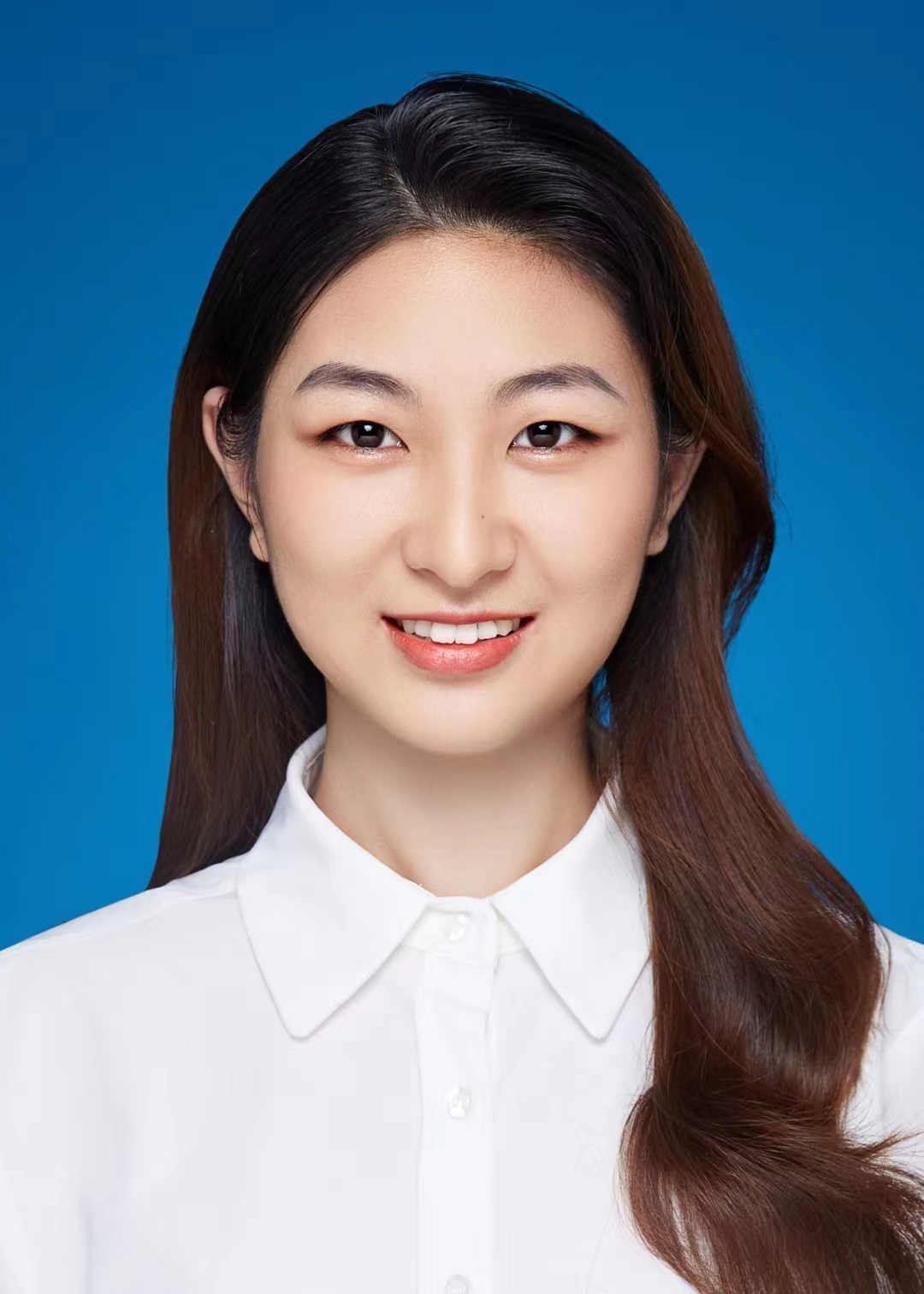}}]{Mengyi Fu} was born in Huaian, Jiangsu, China ,in 2000. She is currently pursuing the Ph.D degree at NanjingUniversity of Posts and Telecommunications. She received her B.Sc from Nanjing University of Posts and Telecommunications, NanjingChina, in 2022. Her research includes encrypted traffic identification, deep learning and traffic prediction.
\end{IEEEbiography}
\begin{IEEEbiography}           [{\includegraphics[height=1.25in,clip,keepaspectratio]{ 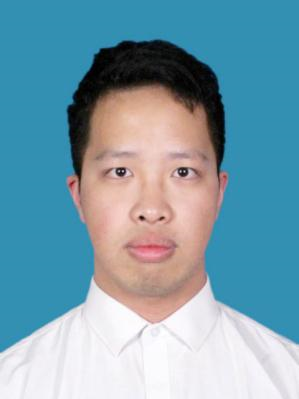}}]{Zixuan Wang} was born in Nanjing, Jiangsu, China ,in 1994 . He obtained a bachelor's degree from Tongda College of Nanjing University of Posts and Telecommunications in 2017, He is currently pursuing a master's degree in logistics engineering at Nanjing University of Posts and Telecommunications. His research interests include encrypted traffic identification and data balancing.
\end{IEEEbiography}
\begin{IEEEbiography}
[{\includegraphics[height=1.25in,clip,keepaspectratio]{ 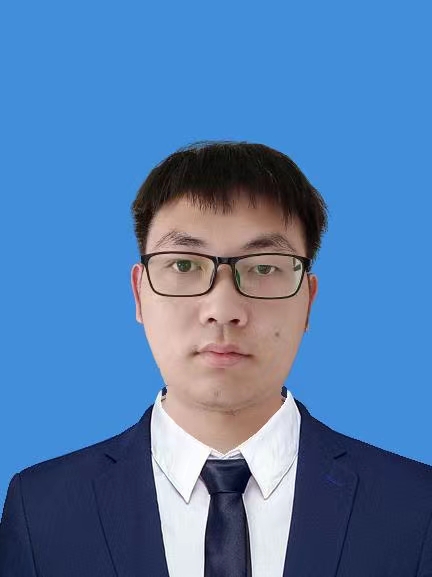}}]
{Ze Zhang} currently pursuing the Master's degree in Information Network at Nanjing University of Postsand Telecommunications. He was born in Zhenjiang, Jiangsu, China, in 2000. He received his Bachelor's degree in Network Engineering in 2022. His research interests include computer networks, network security, anomaly detection and analysis.
\end{IEEEbiography}
\begin{IEEEbiography}
[{\includegraphics[height=1.25in,clip,keepaspectratio]{ 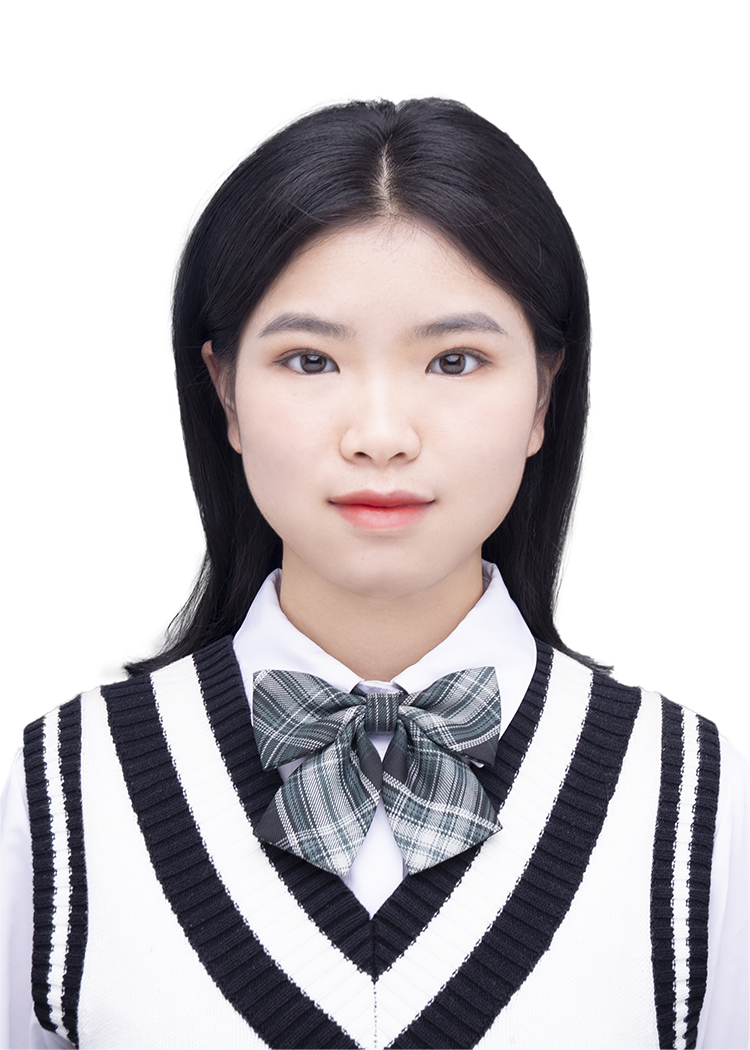}}]
{Minyao Liu} was born in Ganzou, Jiangxi, China, in 2000. She is currently pursuing her master's degree at Nanjing Post and Telecommunications University. She received her bachelor’s degree in Management from NUPT in 2022. Her research areas include traffic identification, deep learning and anomaly detection.
\end{IEEEbiography}

% You can push biographies down or up by placing
% a \vfill before or after them. The appropriate
% use of \vfill depends on what kind of text is
% on the last page and whether or not the columns
% are being equalized.

%\vfill

% Can be used to pull up biographies so that the bottom of the last one
% is flush with the other column.
%\enlargethispage{-5in}

% that's all folks
\end{document}